\documentclass[10pt]{article}
\usepackage{amsmath}
\usepackage{bm}
\usepackage{lscape}
\usepackage{latexsym}
\usepackage[dvips]{graphicx}
\begin{document}
\title{The impacts of asymmetry on modeling and forecasting realized volatility in Japanese stock markets} 
\author{Daiki Maki\thanks{Address: Faculty of Commerce, Doshisha University, 
         Karasuma-higashi-iru, Imadegawa-dori, Kamigyo-ku, Kyoto, Japan 602-8580 
         (E-mail: dmaki@mail.doshisha.ac.jp)}  \\ 
         Doshisha University
         \and Yasushi Ota \thanks{Address: Faculty of Management, Okayama University of Science, 1-1 Ridaicyou, Okayama, Japan 700-0005
         (E-mail: yota@mgt.ous.ac.jp)} \\ 
         Okayama University of Science}
\date{}
\maketitle

\begin{center}
\large{Abstract}
\end{center}
\fontsize{11pt}{25pt}\selectfont
This study investigates the impacts of asymmetry on the modeling and forecasting of realized volatility in the Japanese futures and spot stock markets. 
We employ heterogeneous autoregressive (HAR) models allowing for three types of asymmetry: 
positive and negative realized semivariance (RSV), asymmetric jumps, and leverage effects. 
The estimation results show that leverage effects clearly influence the modeling of realized volatility models. 
Leverage effects exist for both the spot and futures markets in the Nikkei 225. 
Although realized semivariance aids better modeling,   
the estimations of RSV models depend on whether these models have leverage effects. 
Asymmetric jump components do not have a clear influence on realized volatility models.  
While leverage effects and realized semivariance also improve the out-of-sample forecast performance of volatility models,    
asymmetric jumps are not useful for predictive ability. 
The empirical results of this study indicate that asymmetric information, in particular, leverage effects and realized semivariance, yield better modeling and more accurate forecast performance. 
Accordingly, asymmetric information should be included when we model and forecast the realized volatility of Japanese stock markets. 
\\

\newpage
\fontsize{11pt}{27pt}\selectfont
\section{Introduction}
The risk management of asset prices, asset allocation, pricing of derivatives, and hedging depends on the modeling and forecasting of stock market volatility. 
The realized volatility, defined as the sum of the squared return obtained from intraday high-frequency data of asset prices, is significant for the risk management. 
A number of studies have examined the statistical properties of realized volatility over the past two decades.  
The heterogenous autoregressive model (HAR) introduced by Corsi (2009) is one of the most representative realized volatility models. 
The HAR model employs daily, weekly, and monthly information regarding observed realized volatility, and can take into account the long memory properties of volatility. 
Further, the HAR model demonstrates good performance in the forecasting of volatility, despite its simplicity and tractability, and is widely used in the literature. 
In addition, since financial markets occasionally have jumps, 
it is standard for volatility modeling to include jump components (Barndorff-Nielsen and Shephard, 2004; Andersen, Bollerslev, and Diebold, 2007; Corsi, Pirino, and Ren, 2010; 
Sou\v{c}ek and Todorova, 2014; Chen, Ma, and Zhang, 2019).  

The introduction of asymmetry is key to better modeling and forecasting of volatility.
The asymmetry employed in the analysis of realized volatility takes three main forms. 
First is the leverage effect, in which the past negative returns increase future volatility. 
Second is realized semivariance, developed by Barndorff-Nielsen, Kinnebrock, and Shephard (2010), which 
decomposes realized volatility into the parts of positive and negative intraday returns.
Patton and Sheppard (2015) proposed new HAR models based on realized semivariance, and 
referred to the realized volatility of the parts of positive and negative returns as ``good volatility" and ``bad volatility," respectively. 
The third type of asymmetry is asymmetric jump components, which 
reveal the effect of jumps for positive and negative returns on the realized volatility. 
Recent studies have shown that the asymmetric properties of realized volatility have a crucial role in the modeling and forecasting of volatility (Patton and Sheppard, 2015; Audrino and Hu, 2016; Prokopczuk, Symeonidis, and Simen, 2016;
Wen, Gong, and Cai, 2016; Chen et al., 2019). 
However, these studies have not clearly identified what type of asymmetry is most important in the modeling and forecasting of volatility. 
Clearing the problem leads to better modeling and forecasting. 

This study investigates the impacts of asymmetric properties on the modeling and forecasting of realized volatility. 
We particularly focus on the Japanese futures and spot stock markets. 
As described by Bollen and Whaley (2015), the futures price is sensitive to new information, such as financial crises and market changes. 
Since the futures and spot markets are linked to each other, 
it is significant to demonstrate the types of asymmetry that influence volatility modeling and forecasting in these markets. 
The daily amount of trading time in the Japanese futures stock market has varied over the years. 
In addition, the level of foreign investment in the Japanese stock market has been increasing over the last two decades. 
Moreover, the holding period (the period of time that investors hold stocks) has been decreasing.    
It is possible that these facts influence asymmetric properties of realized volatility and the impacts of asymmetry are different with regard to futures and spot markets. 
In this study, we analyze what type of asymmetry is most effective for the modeling and forecasting of volatility, in addition to the differences in the influence of asymmetry on Japanese futures and spot stock markets. 
The results in this study contribute to better modeling and forecasting of realized volatility in Japanese futures and spot stock markets. 

We first demonstrate the impacts of asymmetry on modeling the realized volatility using HAR models with asymmetry. 
We employ the HAR model with a jump proposed by Andersen et al. (2007) as a benchmark model, and expand the models to include leverage effects and asymmetric jumps. 
In addition, we employ and expand a realized semivariance model developed by Patton and Sheppard (2015). 
We estimate these models and then evaluate the out-of-sample forecast performance. 
The empirical results provide evidence that leverage effects have obvious impacts on the modeling of realized volatility models. 
A strong leverage effect exists for both the spot and futures markets in the Nikkei 225, which is a representative stock index in Japan. 
While realized semivariance is useful for better modeling,   
the estimates of realized semivariance models depend on whether these models have leverage effects. 
We also demonstrate that asymmetric jump components do not have clear influences on the realized volatility models.  
Moreover, while leverage effects and realized semivariance also improve the out-of-sample forecast performance,    
asymmetric jumps are not useful for improving predictive ability. 
The comparison in this study indicates that asymmetric information, in particular, leverage effects and realized semivariance, yield better modeling and more accurate forecasting. 
Accordingly, it is necessary to include asymmetric information when we model and forecast realized volatility. 

The remainder of this paper is organized as follows: 
Section 2 presents the theoretical background and empirical models of realized volatility used in this study. 
Section 3 presents empirical results of the estimation of realized volatility models with asymmetry in the Japanese futures and spot markets. 
Section 4 concludes the paper.

\section{Asymmetric realized volatility models}

\subsection{Realized volatility}
We assume that the logarithmic stock price $p_t$ follows the standard jump-diffusion process
\begin{equation}
dp_t=\mu_t dt+\sigma_t dW_t + k_t dq_t, 
\end{equation}
where $\mu_t$ is a drift term with a continuous variation process, $\sigma_t$ is a volatility with a strictly positive stochastic process, 
and $W_t$ is a standard Brownian motion. 
A counting process $q_t$ denotes the jump component such that a jump occurs at $t$ for $dq_t=1$, and does not occur for $dq_t=0$.  
$k_t$ is the size of the jumps. 

The quadratic variation of the log return from $t$ to $t-1$ has 
\begin{equation}
QV_t=[r,r]_t=\int_{t-1}^t \sigma_s^2 ds + \sum_{t-1<s \le t} \kappa_s^2, \ \ \ 0 \le t \le T, 
\end{equation}
where $\int_{t-1}^t \sigma_s^2 ds$ is the integrated variance and 
$\sum_{t-1<s \le t} \kappa_s^2$ denotes the sum of squared jumps between $t-1$ and $t$.
The realized volatility is defined as the sum of squared intraday high-frequency returns: 
\begin{equation}
RV_t=\sum_{j=1}^n r_{t,j}^2, \ \ \ t=1, \cdots, T,
\end{equation}
where $r_{i,j}$ is the intraday return of $j$th $(j=1,\cdots,n)$ on trading day $t$. 
When $n \rightarrow \infty$, the realized volatility is a consistent estimator of $QV_t$:
\begin{equation}
RV_t \stackrel{p}{\to} QV_t.
\end{equation}
$QV_t$ has two components, including the integrated variance and the jump. 
Barndorff-Nielsen and Shephard (2004) show that the bipower variation is a consistent estimator of the integrated variance. 
The bipower variation $BV_t$ is defined as
\begin{equation}
BV_t=\mu_1^{-2} \sum_{j=2}^n |r_{t,j}||r_{t,j-1}| \stackrel{p}{\to} \int_{t-1}^t \sigma_s^2 ds, 
\end{equation}
where $\mu_1=\sqrt{\pi/2}$. 
From the results of (2), (4), and (5), as shown by Andersen et al. (2007), the jump component $J_t= \sum_{t-1<s \le t} \kappa_s^2$ is obtained from 
\begin{equation}
J_t=\max [RV_t-BV_t,0]. 
\end{equation}

\subsection{HAR-type models and asymmetry}
Following the theoretical background of the realized volatility and the heterogeneous market hypothesis, 
Corsi (2009) introduced a simple and tractable heterogeneous autoregressive (HAR) model as an empirical model to demonstrate the dynamics of the realized volatility. 
The HAR model includes daily, weekly, and monthly realized volatilities for explanatory variables. 
This can accommodate the properties of long memory for returns and volatility. 
Andersen et al. (2007) extended it to the model with a jump. 
Based on the HAR models developed by Corsi (2009) and Andersen et al. (2007), 
we use a logarithmic HAR model with a jump as a benchmark. 
As pointed out by Asai, McAleer, and Medeiros (2012b) and Bekierman and Manner (2018), 
the logarithmic model can reduce the influence of heteroscedasticity in the measurement errors when modeling realized volatility.  
For a simplified description of the HAR model, 
we define a vector ${\bf RV_{t-1}}$$=(\ln RV_{t-1}$, $\ln RV_{t-1}^w, \ln RV_{t-1}^m )^{\prime}$, 
where $RV_{t-1}^w=\frac{1}{5}\sum_{i=1}^5 RV_{t-i}$ and  $RV_{t-1}^m=\frac{1}{22}\sum_{i=1}^{22} RV_{t-i}$ are the weekly and monthly averages of the daily realized volatility. 
Based on the study of Andersen et al. (2007), we use $J_t=\max[RV_t-BV_t,0]$ as the daily jump variable and denote the vector as ${\bf J_{t-1}}$$=(\ln (J_{t-1}+1), \ln (J_{t-1}^w+1), \ln (J_{t-1}^m+1))^{\prime}$, 
where $J_{t-1}^w=\frac{1}{5}\sum_{i=1}^5 J_{t-i}$ and $J_{t-1}^m=\frac{1}{22}\sum_{i=1}^{22} J_{t-i}$ are the weekly and monthly averages of the daily jump component. 
The benchmark model is given as 
\begin{equation}
\text{HAR-J:} \ \ \ \ \ \ln RV_t=c+{\boldsymbol  \alpha}^{\prime} {\bf RV_{t-1}} +{\boldsymbol  \beta}^{\prime} {\bf J_{t-1}} +\epsilon_t, 
\end{equation}
where ${\boldsymbol  \alpha}=(\alpha_1,\alpha_2,\alpha_3)^{\prime}$ is the parameter vector of realized volatility variables, 
${\boldsymbol  \beta}=(\beta_1,\beta_2,\beta_3)^{\prime}$ is the parameter vector of jump variables, 
and $\epsilon_t$ is an error term. 

We introduce three types of asymmetry in (7).  
The first is the use of realized semivariance. 
Barndorff-Nielsen et al. (2010) decompose $RV_t$ as upside and downside realized semivariance obtained from positive and negative intraday returns, respectively. 
The positive semivariance estimator is written as
\begin{equation}
RSV_t^{+}=\sum_{j=1}^n r_{t,j}^2 I\{r_{t,j} \ge 0 \},  
\end{equation}
where $I\{\cdot\}$ is an indicator function that takes the value of 1 if $I\{\cdot\}$ is true and 0 if $I\{\cdot\}$ is not true. 
The negative semivariance estimator is written as 
\begin{equation}
RSV_t^{-}=\sum_{j=1}^n r_{t,j}^2 I\{r_{t,j} < 0 \}. 
\end{equation}
From (8) and (9), the daily realized volatility $RV_t$ is decomposed as $RV_t=RSV_t^{+}+RSV_t^{-}$. 
Patton and Sheppard (2015) introduced the HAR model using realized semivariance. 
They referred to upside and downside realized semivariances as ``good volatility" and ``bad volatility," respectively. 
While they use only daily components, Todorova (2017) extends the model by adding weekly and monthly realized semivariance components. 
However, Todorova (2017) does not add other components, including jumps and leverage effects. 
We generalize and propose the model$^{1}$.  
We denote the vector of realized semivariance as ${\bf RSV_{t-1}}$$=(\ln RSV_{t-1}^{+}$, $\ln RSV_{t-1}^{w+}, \ln RSV_{t-1}^{m+}, \ln RSV_{t-1}^{-}$, $\ln RSV_{t-1}^{w-}, \ln RSV_{t-1}^{m-}, )^{\prime}$, 
where $RSV_{t-1}^{w+}=\frac{1}{5}\sum_{t-i}^5 RSV_{t-i}^{+}$ and $RSV_{t-1}^{w-}=\frac{1}{5}\sum_{t-i}^5 RSV_{t-i}^{-}$ are weekly averages of positive and negative semivariances 
and $RSV_{t-1}^{m+}=\frac{1}{22}\sum_{t-i}^{22} RSV_{t-i}^{+}$ and $RSV_{t-1}^{m-}=\frac{1}{22}\sum_{t-i}^{22} RSV_{t-i}^{+}$ are monthly averages of positive and negative semivariances. 

The realized semivariance model with the jump is expressed as 
\begin{equation}
\text{RSV-J:} \ \ \ \ \ \ln RV_t=c+{\boldsymbol  \tilde{\alpha}}^{\prime} {\bf RSV_{t-1}} +{\boldsymbol  \beta}^{\prime} {\bf J_{t-1}} +\epsilon_t, 
\end{equation}
where ${\boldsymbol  \tilde{\alpha}}=(\alpha_1, \alpha_2, \alpha_3, \alpha_4, \alpha_5, \alpha_6)^{\prime}$ is the parameter vector for realized semivariance variables. 

The second asymmetry is for jump variables. 
Barndorff-Nielsen et al. (2010) also show that positive semivariance (8) and negative semivariance (9) converge to one-half of the integrated variance, with positive or negative jump components as follows: 
\begin{gather}
RSV_t^{+}  \stackrel{p}{\to} \frac{1}{2}\int_{t-1}^t \sigma_s^2 ds + \sum_{t-1<s \le t} \kappa_s^2 I\{\kappa_s \ge 0 \}, \\
RSV_t^{-}  \stackrel{p}{\to} \frac{1}{2}\int_{t-1}^t \sigma_s^2 ds + \sum_{t-1<s \le t} \kappa_s^2 I\{\kappa_s < 0 \},  
\end{gather}
where $\sum_{t-1<s \le t} \kappa_s^2 I\{\kappa_s \ge 0 \}$ and $\sum_{t-1<s \le t} \kappa_s^2 I\{\kappa_s < 0 \}$ are positive and negative jump components, respectively. 
Using the properties (5), (11), and (12), we can obtain positive and negative jump components given by
\begin{gather}
J_t^{+}=\max [RSV_t^{+}-\frac{1}{2}BV_t,0], \\
J_t^{+}=\max [RSV_t^{-}-\frac{1}{2}BV_t,0].
\end{gather}
We denote the vector of asymmetric jump variables as ${\bf AJ_{t-1}}$$=(\ln (J_{t-1}^{+}+1), \ln (J_{t-1}^{w+}+1), \ln (J_{t-1}^{m+}+1), \ln (J_{t-1}^{-}+1), \ln (J_{t-1}^{w-}+1), \ln (J_{t-1}^{m-}+1))^{\prime}$, 
where $J_{t-1}^{w+}=\frac{1}{5}\sum_{i=1}^5 J_{t-i}^{+}$ and $J_{t-1}^{w-}=\frac{1}{5}\sum_{i=1}^5 J_{t-i}^{-}$ are weekly averages of daily positive and negative jump components, 
and $J_{t-1}^{m+}=\frac{1}{22}\sum_{i=1}^{22} J_{t-i}^{+}$ and $J_{t-1}^{m-}=\frac{1}{22}\sum_{i=1}^{22} J_{t-i}^{-}$ are monthly averages of daily positive and negative jump components. 
The HAR model with asymmetric jumps is written as 
\begin{equation}
\text{HAR-AJ:} \ \ \ \ \ \ln RV_t=c+{\boldsymbol  \alpha}^{\prime} {\bf RV_{t-1}} +{\boldsymbol  \tilde{\beta}}^{\prime} {\bf AJ_{t-1}} + \epsilon_t, 
\end{equation}
where ${\boldsymbol  \tilde{\beta}}=(\beta_1, \beta_2, \beta_3, \beta_4, \beta_5, \beta_6)^{\prime}$ is the parameter vector for asymmetric jump variables. 
Patton and Sheppard (2015) show that this formulation can measure the effects of positive and negative jump components on the realized volatility$^2$.

Similarly, we also use the realized semivariance model with the asymmetric jump components given by
\begin{equation}
\text{RSV-AJ:} \ \ \ \ \ \ln RV_t=c+{\boldsymbol  \tilde{\alpha}}^{\prime} {\bf RSV_{t-1}} +{\boldsymbol  \tilde{\beta}}^{\prime} {\bf AJ_{t-1}} + \epsilon_t. 
\end{equation}
Equation (16) has two types of asymmetry, including realized volatility and jump components. 

The third asymmetry is to introduce the leverage effect. 
The leverage effect generally refers to the negative relationship between the asset return and the change of volatility.  
Bollerslev, Litvinova, and Tauchen (2006) and A\"{i}t-Sahalia, Fan, and Li (2013) investigate the leverage effect in high-frequency data. 
Asai, McAleer, and Medeiros (2012a), Corsi and Ren\`{o} (2012), and Gong and Lin (2019) show that the forecasting performance of realized volatility is improved by introducing a persistent leverage effect$^{3}$. 
We define the return vector as ${\bf r_{t-1}}$$=(|r_{t-1}|,$ $|r_{t-1}^w|$, $|r_{t-1}^m|$,$|r_{t-1}|I\{r_{t-1}<0\}$, $|r_{t-1}^w|I\{r_{t-1}^w<0\}$, $|r_{t-1}^m|I\{r_{t-1}^m<0\})^{\prime}$, 
where $|r_{t-1}^w|=\frac{1}{5}\sum_{i=1}^5 |r_{t-i}|$ and  $|r_{t-1}^m|=\frac{1}{22}\sum_{i=1}^{22} |r_{t-i}|$ are weekly and monthly average (absolute) returns. 
While absolute returns $|r_{t-1}|$, $|r_{t-1}^w|$, and $|r_{t-1}^m|$ govern the size and effect of the return, 
$|r_{t-1}|I\{r_{t-1}<0\}$, $|r_{t-1}^w|I\{r_{t-1}^w<0\}$, and $|r_{t-1}^m|I\{r_{t-1}^m<0\}$ are variables for the leverage effect.
Horpestad, Ly\'{o}csa, Moln\'{a}r, and Olsen  (2019) employ this type of model and use only the daily absolute return $|r_{t-1}|$ and leverage effect about $|r_{t-1}|I\{r_{t-1}<0\}$. 
We develop the models for the long memory return and leverage effect by adding weekly and monthly average variables. 
The HAR-based model with leverage effect is given as 
\begin{equation}
\text{HAR-J-LE:} \ \ \ \ \ \ln RV_t=c+{\boldsymbol  \alpha}^{\prime} {\bf RV_{t-1}} +{\boldsymbol  \beta}^{\prime} {\bf J_{t-1}} +{\boldsymbol  \delta}^{\prime} {\bf r_{t-1}} + \epsilon_t, 
\end{equation}
where ${\boldsymbol  \delta}=(\delta_1, \delta_2, \delta_3, \delta_4, \delta_5, \delta_6)^{\prime}$ is the parameter vector of return variables. 
When $\delta_1>0$ and $\delta_4>0$ or $\delta_4 >0$ and $\delta_4>\delta_1$, the clear (daily) leverage effect exists. 
When $\delta_4=0$ and $\delta_1>0$, the realized volatility depends on only the absolute scale of the daily return.  

The realized semivariance model with the leverage effect is written as
\begin{equation}
\text{RSV-J-LE:} \ \ \ \ \ \ln RV_t=c+{\boldsymbol \tilde{\alpha}}^{\prime} {\bf RSV_{t-1}} +{\boldsymbol  \beta}^{\prime} {\bf J_{t-1}} +{\boldsymbol  \delta}^{\prime} {\bf r_{t-1}} + \epsilon_t.  
\end{equation}
Equation (18) has two types of asymmetric properties of realized semivariance and leverage effects. 

From (15) and (17), we can develop the HAR model with asymmetric jump components and leverage effects. 
\begin{equation}
\text{HAR-AJ-LE:} \ \ \ \ \ \ln RV_t=c+{\boldsymbol  \alpha}^{\prime} {\bf RV_{t-1}} +{\boldsymbol \tilde{\beta}}^{\prime} {\bf AJ_{t-1}} + {\boldsymbol  \delta}^{\prime} {\bf r_{t-1}} +\epsilon_t,
\end{equation}
Equation (19) takes two asymmetric properties of jump components and leverage effects.

Similarly, we can introduce the RSV model with both asymmetric jump components and leverage effects as follows: 
\begin{equation}
\text{RSV-AJ-LE:} \ \ \ \ \ \ln RV_t=c+{\boldsymbol  \tilde{\alpha}}^{\prime} {\bf RSV_{t-1}} + {\boldsymbol  \tilde{\beta}}^{\prime} {\bf AJ_{t-1}}+{\boldsymbol  \delta}^{\prime} {\bf r_{t-1}} +\epsilon_t. 
\end{equation}
Equation (20) has three types of asymmetric properties, including realized semivariance, jump components, and leverage effects.

\section{Empirical analysis}

\subsection{Data and statistics}
This section investigates the impacts of asymmetric properties on modeling and forecasting of realized volatility of futures and spot stock markets of the Nikkei 225, a major Japanese stock index. 
We use 5-min high-frequency data 
because, as pointed out by Andersen et al. (2007) and Corsi et al (2010), 5-min high-frequency data mitigate the impact of the market microstructure noise on estimates 
and provide more accurate measurement. 
The sample period used in the study is from January 4, 2001, to September 20, 2019. 
The daily amount of trading time of the Japanese futures stock market has varied over the years. 
The evening trading session first began on September 18, 2007, and the trading time of the evening session was first extended on October 14, 2008. 
The intraday data from January 4, 2001, to September 18, 2007, have $n=58$, before the introduction of the evening session. 
We have intraday data $n=89$ from September 9, 2007, to October 14, 2008, $n=101$ from October 15, 2008, to July 20, 2010,  
and $n=143$ from July 21, 2010, to July 15, 2011, due to the introduction of the evening session and the extension of the trading time. 
Intraday data from July 19, 2011, to July 18, 2016, and from July 20, 2016, to September 20, 2019, have $n=203$ and $n=236$, 
because the night session had begun and the trading system had changed. 
The high-frequency data from the Japanese futures stock market of Nikkei 225 are obtained from 225Labo$^{4}$.
The realized volatility data of the Japanese spot stock market of Nikkei 225 were obtained from Oxford-Man Institute of Quantitative Finance$^{5}$. 
The futures market and spot market data include 4,568 and 4,552 days, respectively. 

Tables 1 and 2 contain the descriptive statistics of realized volatility, jump, and return variables for the futures and spot markets of Nikkei 225. 
Generally, the basic numbers for the futures market are larger than those for the spot market. 
For example, the mean, maximum, and standard deviation of $RV_t$ for the futures market are 0.2432, 32.995, and 0.6934, 
whereas those for the spot market are 0.1005, 3.2288, and 0.1647. 
Figures 1, 2, 3, and 4, which draw $RV_t$ and $\ln RV_t$ for futures and spot markets, also indicate larger numbers for the futures market than the spot market.  
It is also observed that the realized volatility has jumps and autocorrelation from all figures.  
In fact, the Ljung-Box $Q$ statistics show that the null hypothesis of no autocorrelation up to 20th order is rejected for all variables at 1\% significance. 
The results imply that weekly and monthly averages for all variables are useful to model and forecast realized volatility. 
Comparing $Q$ statistics of the futures market and spot market, while those of minus variables ($RV_t^{-}$) for  the futures market are larger than those of plus variables, 
those of minus variables ($RV_t^{-}$) for the spot market are smaller than those of plus variables. 

\subsection{Estimations of each model}
Table 3 tabulates the estimation results of regression models (7), (10), (15), (16), (17), (18), (19), and (20) for the futures market of Nikkei 225 over the in-sample period. 
We use the Newey-West estimator to correct autocorrelation and heteroscedasticity of the error term$^{6}$. 
From the estimation results of HAR models, we see that the estimates of $\alpha_1$, $\alpha_2$, and $\alpha_3$ are significant at 1\% level. 
This indicates that $RV_{t-1}$, $RV_{t-1}^w$, and $RV_{t-1}^m$ are useful for the forecast of the future $RV_t$.  
Jump components have different effects on each HAR model. 
Only estimates of $\beta_2$ for HAR-J and $\beta_3$ for HAR-AJ are significant at 10\% or 5\% levels. 
Estimates of other jump parameters are not significant. 
The impacts of jump components are small for HAR-J and HAR-AJ.  
In contrast, HAR-J-LE and HAR-AJ-LE have clear effects of jump components on $RV_t$. 
For HAR-J-LE, past jumps decrease future RV. 
For HAR-AJ-L, weekly and monthly positive jump components increase $RV_t$, 
whereas weekly and monthly negative jump components decrease $RV_t$. 
This implies an asymmetric effect of jump components on realized volatility. 

It should be noted that HAR-J-LE and HAR-AJ-LE have leverage effects on the model.
The results show that the estimates and significance of jump components depend on whether the regression model includes leverage effects.   
We also find that leverage effects for HAR-J-LE and HAR-AJ-L are clear.   
Estimates of $\delta_4$, $\delta_5$, and $\delta_6$ are significant at 1\% or 10\% levels for HAR-J-LE and 1\% levels for HAR-AJ-LE, 
indicating that the past negative return is useful for forecasting $RV_t$. 
In addition, the differences of Adj.$R^2$ between HAR-J and HAR-AJ and between HAR-J-LE and HAR-AJ-LE are small. 
However, the differences of Adj.$R^2$ between HAR-J and HAR-J-LE and between HAR-AJ and HAR-AJ-LE are obvious. 
The results provide evidence that the leverage effect clearly influences the modeling of realized volatility.  

The use of positive and negative semivariance brings higher Adj.$R^2$ than HAR models. 
For example, Adj.$R^2$ of HAR-J is 0.564, whereas that of RSV-J is 0.590. 
The realized semivariances are effective for better models. 
The estimation results of RSV models also exhibit asymmetric effects of RSV, jump, and return variables on realized volatility. 
For RSV-J and RSV-AJ, estimates of negative semivariance parameters $\alpha_4$ and $\alpha_5$ are larger than those of positive semivariance parameters $\alpha_1$ and $\alpha_2$.  
This indicates that negative semivariances increase the future realized volatility compared to positive semivariances. 
Similar to the results of  HAR-J and HAR-AJ, the impacts of jump components are small for RSV-J and RSV-AJ.  
In contrast, we find that unlike HAR models, jump components of RSV-J-LE and RSV-AJ-LE do not perform well. 
Further, the differences of estimates of positive and negative semivariance parameters between RSV-J-LE and RSV-AJ-LE models are smaller than those of between RSV-J and RSV-AJ. 
RSV-AJ has estimates $(\hat{\alpha}_1,\hat{\alpha}_4)=(0.087, 0.243)$ for $RSV_{t-1}^{+}$ and $RSV_{t-1}^{-}$, and $(\hat{\alpha}_2,\hat{\alpha}_5)=(0.077, 0.266)$ for $RSV_{t-1}^{w+}$ and $RSV_{t-1}^{w-}$, 
whereas RSV-AJ-LE has $(\hat{\alpha}_1,\hat{\alpha}_4)=(0.106, 0.134)$ and $(\hat{\alpha}_2,\hat{\alpha}_5)=(0.158, 0.144)$. 
It is possible that the different estimation results are affected by the presence of leverage effects.

Table 4 contains estimation results of the spot market of Nikkei 225. 
We can generally observe properties similar to those of Table 3.  
Most estimates of realized volatility and leverage effects parameters are significant.  
In addition, the presence of leverage effects has impacts on estimations. 
However, some differences are observed between Tables 3 and 4. 
The estimates of $\alpha_1$ for HAR models in Table 3 are smaller than those of $\alpha_2$. 
This indicates that the influence of the weekly realized volatility variable is larger than that of the daily realized volatility variable.  
In contrast, the estimates of $\alpha_1$ for HAR models in Table 4 are similar to those of $\alpha_2$. 
Past daily variables are more important to explain and model the realized volatility in the spot market.   
Unlike the estimations of RSV-J-LE and RSV-AJ-LE in Table 3, 
the impacts of negative semivariance on the future volatility in Table 4 is larger, even if the regression includes leverage effects. 
For example, RSV-AJ-LE has the estimates $(\hat{\alpha}_2,\hat{\alpha}_4)=(0.158, 0.144)$ in Table 3 and $(\hat{\alpha}_2,\hat{\alpha}_4)=(0.139, 0.212)$ in Table 4. 
The estimate of the daily scale return effect parameter $\delta_1$ is significant and the monthly leverage effect $\delta_6$ is not significant in Table 4. 
For the spot market, the latest information has larger effects on modeling realized volatility. 

In-sample analysis provides some findings of the impact of asymmetry on modeling the realized volatility. 
The estimation results show that the leverage effect clearly influences the modeling of realized volatility models. 
The leverage effect clearly exists for both HAR and RSV models for the spot and futures markets of Nikkei 225. 
In particular, the daily and weekly effects are stronger. 
The presence of the leverage effect also increases Adj.$R^2$ for HAR and RSV models. 
While realized semivariances are useful for better modeling,   
the estimations of realized semivariance parameters for RSV models depend on whether RSV models have the leverage effect. 
Asymmetric jump components do not have clear influences on the estimates of realized volatility variables for HAR and RSV models.

\subsection{Forecast performance}
We next compare the out-of-sample forecast performance to evaluate the impact of asymmetry on prediction. 
The standard rolling window method is employed. 
The in-sample window contains 1,000 days. 
The estimation period is rolled forward by adding one new day and dropping the oldest day. 
First, we estimate each regression model from the 1st to 1,000th periods, and obtain the out-of-sample predicted values at the 1,001st period. 
Next, we move the window forward 1 day and estimate each regression model from the 2nd to 1,001st periods. 
The out-of-sample predicted values at the 1,002nd period are obtained from the method.  
Thus, we estimate regression models for each sample period and obtain the out-of-sample predicted values of 1-day volatility $\ln RV_t$. 

To evaluate the out-of-sample forecast performance, 
we use four loss functions, including the mean squared error (MSE), the mean absolute error (MAE), the heteroscedasticity adjusted mean squared error (HMSE), and the heteroscedasticity adjusted mean absolute error (HMAE). 
They are defined as follows:
\begin{gather}
MSE=M^{-1}\sum_{t=1}^M(\ln \hat{RV}_t- \ln RV_t)^2, \\ 
MAE=M^{-1}\sum_{t=1}^M|\ln \hat{RV}_t- \ln RV_t|, \\
HMSE=M^{-1}\sum_{t=1}^M(1-\ln \hat{RV}_t/ \ln RV_t)^2, \\ 
HMAE=M^{-1}\sum_{t=1}^M|1-\ln \hat{RV}_t/ \ln RV_t|, 
\end{gather}
where $M$ is the number of forecast days and $\ln \hat{RV}_t$ denotes the out-of-sample predicted values of each model. 
Based on the four loss functions, we use the DM test introduced by Diebold and Mariano (1995) and compare the forecast performance. 
The null hypothesis of the DM test is that the forecast performance of two models has the same accuracy. 
The alternative hypothesis is that the forecast performance of two models is different. 
The DM test statistics have the limiting distribution with $N(0,1)$. 

Tables 5 and 6 report the results of DM tests for the futures and spot markets of Nikkei 225. 
To examine the impacts of asymmetric properties on prediction, 
we compare all combinations of benchmark and comparison models.
When the benchmark model is HAR-J in Table 5, 
the DM statistic based on MAE for the comparison model HAR-J-LE is 8.445 and the null hypothesis is rejected at 1\% significance levels.
The result indicates that HAR-J-LE has superior forecast performance to HAR-J. 
Thus, the regression model with the leverage effect brings better predictive ability. 
We can see that all comparison models perform better than the benchmark model HAR-J in Table 6. 
This implies that asymmetric information is effective for the out-of-sample forecast in the spot market. 
The DM statistic based on MAE and HMAE for the comparison model RSV-J is 6.361 in Table 5 and 3.966 in Table 6, and is significant at 1\% level.
The forecast performance of realized semivariance model RSV-J is better than that of HAR-J. 
By contrast, the DM statistics of the difference between the benchmark HAR-J and comparison HAR-AJ are not significant in Tables 5. 
Asymmetric jump components do not have better impacts on the out-of-sample forecast in the futures market. 
In fact, all comparison models have better predictive ability in Table 5 when HAR-AJ is the benchmark model. 
 
We find that the HAME based DM statistic of the difference between the benchmark HAR-J-LE and comparison RSV-J is -4.917, and the null hypothesis is rejected at 1\% level in Table 5. 
The statistic in Table 6 is -2.343 and 5\% significance level. 
This shows that the benchmark model performs better than the comparison model. 
However, the comparison model RSV-J-LE outperforms the benchmark model HAR-J-LE. 
RSV-J-LE has dominated forecast performance except for RSV-ASJ-L, because all DM statistics based on MAE and HMAE are significant at 1\% or 5\% levels in Table 5. 
When the benchmark model is RSV-J-LE and the comparison model is RSV-AJ-LE, the DM test does not reject the null hypothesis. 
While the leverage effect and realized semivariance increase the out-of-sample forecast performance,    
asymmetric jumps are not useful for predictive ability. 
In Table 6, HAR-J-LE, HAR-AJ-LE, RSV-J-LE, and RSV-AJ-LE have similar forecast performance.

\section{Summary and conclusion}
The introduction of asymmetry is key to better modeling and forecasting of volatility. 
Although recent studies have shown that asymmetric properties of the realized volatility have a crucial role in the modeling and forecasting of volatility, 
it is unclear what type of asymmetry is the most important.  
This study investigates the impacts of asymmetry on the modeling and forecasting of realized volatility in the Japanese futures and spot stock markets. 
We employed heterogeneous autoregressive models (HAR) allowing for three types of asymmetry: 
positive and negative realized semivariance (RSV), asymmetric jump (AJ), and leverage effect (LE). 
We compared eight realized volatility models, including those of HAR-J, HAR-AJ, HAR-J-LE, HAR-AJ-LE, RSV-J, RSV-AJ, RSV-J-LE, and RSV-AJ-LE. 
The estimation results show that the leverage effect clearly exists for both the spot and futures markets of Nikkei 225. 
The leverage effect clearly influences the modeling of RV models. 
Although realized semivariance is useful for better modeling,   
the estimations of RSV models depend on whether RSV models have the leverage effect. 
Asymmetric jump components do not have clear impacts on realized volatility models.  
Moreover, while leverage effects and realized semivariances also increase the out-of-sample forecast performance,    
asymmetric jumps are not useful for improving predictive ability. 
The comparison in this study indicates that asymmetric information, in particular, leverage effects and realized semivariances, yield better modeling and more accurate forecast performance. 
Accordingly, it is necessary to include asymmetric information when we model and forecast realized volatility. 
Empirical evidence, such as stock indexes in other countries, commodities markets, and the impacts of asymmetry on volatility spillover, lead to better modeling and forecasting of realized volatility 
and the appropriate risk management. 
They are left for further study.

\newpage
\fontsize{10pt}{20pt}\selectfont
\begin{flushleft}
{\Large Footnotes}\\
\end{flushleft} 
1. Wen et al. (2016) introduced some extended realized semivariance models with structural breaks. 
\\
\\
2. Patton and Sheppard (2015) also propose another jump variable, called the signed jump variable, to consider the asymmetric jump effects.       
Wen et al. (2016) employ weekly and monthly averages of signed jumps for modeling realized volatility. 
Prokopczuk et al. (2016) and Tauchen and Zhou (2011) propose other asymmetric jump components. 
\\ 
\\
3. Qu, Chen, Niu, and Li (2016), Ma, Wahab, Huang, and Xu (2017), and Qu, Duan, and Niu (2018) develop other asymmetric models with leverage effects. 
\\ 
\\ 
4. URL is https://225labo.com/.  
\\ 
\\
5. URL is  https://realized.oxford-man.ox.ac.uk/data/download. 
\\
\\
6. The lag specification of the Newey-West estimator is determined by Akaike Information Criterion (AIC). 
The Bartlett kernel with the Newey-West automatic bandwidth selection is employed. 

\newpage
\fontsize{11pt}{19pt}\selectfont

\newpage
\fontsize{11pt}{15pt}\selectfont
\begin{landscape}
\begin{center}
Table 1: Descriptive statistics of realized volatility, jump, and return variables for the futures market of Nikkei 225
\end{center}

\begin{center}
\begin{tabular}{ccccccccc} \hline       
                           &Mean          &Median     &Maximum     & Minimum    &Std. Dev.            &Skewness        &Kurtosis        & Q(20)                                      \\ \hline 
$RV_t$                  &0.2432         &0.1439     &32.995         &0.0087       &0.6934               &29.840            &1229.3          & 5255.3$^{\ast \ast \ast}$           \\
$RV_t^{+}$              &0.1213         &0.0652     &32.304         &0.0040       &0.5225               &51.950            &3157.4          & 833.19$^{\ast \ast \ast}$            \\
$RV_t^{-}$              &0.1218         &0.0642     &16.699         &0.0040       &0.3850               &28.251            &1075.2          & 3227.7$^{\ast \ast \ast}$           \\
$J_t$                    &0.1331         &0.0437     &31.671         &0.0000        &0.6450               &32.449            &1398.0          & 2699.8$^{\ast \ast \ast}$           \\
$J_t^{+}$                &0.0531         &0.0123     &31.390         &0.0000        &0.4881               &58.385            &3724.8          & 207.75$^{\ast \ast \ast}$            \\
$J_t^{-}$               &0.0528          &0.0113     &15.142         &0.0000        &0.3283               &34.567           &1454.3            & 1460.7$^{\ast \ast \ast}$            \\
$r_t$                     &0.0001         &0.0008     &0.2136         &-0.1019     &0.0153               &0.2115            &12.632           & 64.466$^{\ast \ast \ast}$            \\
$\ln RV_t$             &-8.8027       &-8.8459    &-3.4113       &-11.650      &0.8415               &0.6639            &1.6060         &  25431$^{\ast \ast \ast}$             \\
$\ln RV_t^{+}$         &-9.5637       &-9.6376   &-3.4325        &-12.416     &0.8799               &0.7154            &1.5413          &  16425$^{\ast \ast \ast}$             \\
$\ln RV_t^{-}$         &-9.6018       &-9.6526   &-4.0923        &-12.424     &0.9297               &0.6507            &1.3601          &   18136$^{\ast \ast \ast}$             \\
$\ln (J_t+1)$          &0.1329         &0.0437     &31.180         &0.0000        &0.6383               &32.119            &1373.6          &  2741.4$^{\ast \ast \ast}$            \\
$\ln (J_t^{+}+1)$      &0.0530         &0.0123     &30.907         &0.0000        &0.4812               &58.154            &3703.8          &  213.58$^{\ast \ast \ast}$            \\
$\ln (J_t^{-}+1)$     &0.0527         &0.0077     &15.028          &0.0000       &0.3264               &34.439           &1446.0          &  1464.4$^{\ast \ast \ast}$            \\
$|r_t|$                    &0.0106         &0.0077     &0.2136         &0.0000        &0.0110               &3.6937            &35.000          &  4200.6$^{\ast \ast \ast}$            \\
$|r_t|I_t{ \{r_t<0\} }$  &0.0052         &0.0000     &0.1019         &0.0000         &0.0096               &3.2983            &16.935          &  834.84$^{\ast \ast \ast}$            \\        
    \hline 
   \end{tabular}
\end{center}
$RV_t$, $RV_t^{+}$, $RV_t^{-}$, $J_t$, $J_t^{+}$, $J_t^{-}$, $\ln (J_t+1)$, $\ln (J_t^{+}+1)$, $\ln (J_t^{-}+1)$ in Table 1 are multiplied by 1,000. 
$Q(20)$ is the Ljung-Box statistic for up to 20th order serial correlation. The results of other order serial correlations are available upon request. 
$^{**}$ and $^{*}$  denote rejections of null hypothesis at 1\% and 5\% significance levels, respectively.
\end{landscape}

\newpage
\fontsize{11pt}{15pt}\selectfont
\begin{landscape}
\begin{center}
Table 2: Descriptive statistics of realized volatility, jump, and return variables for the spot market of Nikkei 225
\end{center}

\begin{center}
\begin{tabular}{ccccccccc} \hline       
                           &Mean          &Median     &Maximum     & Minimum    &Std. Dev.            &Skewness        &Kurtosis        & Q(20)           \\ \hline 
$RV_t$                  &0.1005         &0.0595     &3.2288         &0.0039       &0.1647               &8.5346            &107.94          & 16221$^{\ast \ast \ast}$              \\
$RV_t^{+}$              &0.0487         &0.0278     &1.5963         &0.0016       &0.0812               &8.1414            &103.93          & 14585$^{\ast \ast \ast}$              \\
$RV_t^{-}$              &0.0517         &0.0276     &2.2322         &0.0013       &0.1040               &10.650            &161.98          & 8222.5$^{\ast \ast \ast}$              \\
$J_t$                    &0.0146         &0.0062     &1.0589         &0.0000        &0.0322               &14.520            &372.11          & 508.01$^{\ast \ast \ast}$              \\
$J_t^{+}$               &0.0107         &0.0021     &0.5269        &0.0000        &0.0268               &7.3791            &85.384          & 1242.1$^{\ast \ast \ast}$              \\
$J_t^{-}$               &0.0126         &0.0021     &1.2651         &0.0000      &0.0149               &15.209              &353.59          & 1088.9$^{\ast \ast \ast}$             \\
$r_t$                     &0.0001         &0.0005     &0.1323         &-0.1211     &0.0149               &-0.4271            &6.5450          &526.86$^{\ast \ast \ast}$             \\
$\ln RV_t$             &-9.6996       &-9.7293    &-5.7356       &-12.441      &0.9308               &0.3270            &0.8664         &  33828$^{\ast \ast \ast}$               \\
$\ln RV_t^{+}$         &-10.451       &-10.489   &-6.4400        &-13.313     &0.9596              &0.3293            &0.9210          &   26728$^{\ast \ast \ast}$               \\
$\ln RV_t^{-}$         &-10.461       &-10.494   &-6.1047        &-13.504     &1.0185               &0.2740            &1.0375          &  24821$^{\ast \ast \ast}$                \\
$\ln (J_t+1)$          &0.0146         &0.0062     &1.0583         &0.0000        &0.0322               &14.512            &371.76          & 508.39$^{\ast \ast \ast}$               \\
$\ln (J_t^{+}+1)$     &0.0107         &0.0021     &0.5267         &0.0000        &0.0268               &7.3778            &85.351          & 1089.0$^{\ast \ast \ast}$               \\
$\ln (J_t^{-}+1)$     &0.0126        &0.0021      &1.2643         &0.0000       &0.0416               &15.202             &353.26          & 527.29$^{\ast \ast \ast}$               \\
$|r_t|$                    &0.0105         &0.0075     &0.1323         &0.0000        &0.0106              &2.8603            &0.0001          &  3645.4$^{\ast \ast \ast}$               \\
$|r_t|I_t{ \{r_t<0\} }$  &0.0052         &0.0000     &0.1211         &0.0000         &0.0096               &3.5661            &0.0001         &  684.50$^{\ast \ast \ast}$              \\        
    \hline 
   \end{tabular}
\end{center}
$RV_t$, $RV_t^{+}$, $RV_t^{-}$, $J_t$, $J_t^{+}$, $J_t^{-}$, $\ln (J_t+1)$, $\ln (J_t^{+}+1)$, $\ln (J_t^{-}+1)$ in Table 1 are multiplied by 1,000. 
$Q(20)$ is the Ljung-Box statistic for up to 20th order serial correlation. The results of other order serial correlations are available upon request. 
$^{***}$, $^{**}$, and $^{*}$ denote rejections of null hypothesis at 1\%, 5\%, and 10\% significance levels, respectively.
\end{landscape}

\newpage
\fontsize{11pt}{14pt}\selectfont
\begin{landscape}
\begin{center}
Table 3: Estimation results for the futures market of Nikkei 225
\end{center}

\begin{center}
\begin{tabular}{c|cccc|cccc} \hline       
                 &HAR-J                         &HAR-AJ                    &HAR-J-LE                     & HAR-AJ-LE                 &RSV-J                          &RSV-AJ                   &RSV-J-LE                       & RSV-AJ-LE           \\ \hline 
constant     &                                   &                                  &                                  &                                   &                                   &                                 &                                   &                              \\
$c$            &-0.887$^{\ast \ast \ast}$&-1.204$^{\ast \ast \ast}$&-1.891$^{\ast \ast \ast}$&-2.015$^{\ast \ast \ast}$&-0.133                         &-0.407$^{\ast \ast \ast}$&-1.123$^{\ast \ast \ast}$&-1.300$^{\ast \ast \ast}$\\
RV parameters&           &           &           &            &            &            &              &                 \\
$\alpha_1$  &0.274$^{\ast \ast \ast}$ &0.278$^{\ast \ast \ast}$&0.167$^{\ast \ast \ast}$   &0.167$^{\ast \ast \ast}$   &0.083$^{\ast \ast \ast}$  &0.087$^{\ast \ast \ast}$& 0.104$^{\ast \ast \ast}$& 0.106$^{\ast \ast \ast}$\\
$\alpha_2$  &0.365$^{\ast \ast \ast}$ &0.349$^{\ast \ast \ast}$&0.316$^{\ast \ast \ast}$   &0.315$^{\ast \ast \ast}$   &0.070$^{\ast \ast \ast}$  &0.077$^{\ast \ast \ast}$& 0.160$^{\ast \ast \ast}$ &0.158$^{\ast \ast \ast}$\\
$\alpha_3$  &0.266$^{\ast \ast \ast}$ &0.244$^{\ast \ast \ast}$&0.325$^{\ast \ast \ast}$   &0.312$^{\ast \ast \ast}$   &0.132$^{\ast \ast \ast}$  &0.166$^{\ast \ast \ast}$& 0.234$^{\ast \ast \ast}$&0.153$^{\ast \ast \ast}$\\  
$\alpha_4$  &                                 &                                  &                                   &                                    &0.238$^{\ast \ast \ast}$  &0.243$^{\ast \ast \ast}$& 0.136$^{\ast \ast \ast}$ &0.134$^{\ast \ast \ast}$\\
$\alpha_5$  &                                 &                                  &                                   &                                     &0.272$^{\ast \ast \ast}$  &0.266$^{\ast \ast \ast}$& 0.146$^{\ast \ast \ast}$& 0.144$^{\ast \ast \ast}$\\
$\alpha_6$  &                                 &                                  &                                   &                                    &0.117$^{\ast \ast \ast}$   &0.045                         & 0.044                         & 0.111$^{\ast \ast }$       \\  
Jump parameters&                           &           &           &            &            &            &              &                 \\
$\beta_1$   &4.591                          &-51.14                        &-42.34$^{\ast \ast \ast}$   &-61.00$^{\ast \ast \ast}$  &6.421                            &-20.56                      &-49.43$^{\ast \ast}$      &-56.05$^{\ast \ast}$      \\
$\beta_2$   &-38.00$^{\ast}$             &-15.90                        &-29.11$^{\ast \ast \ast}$   &92.58$^{\ast \ast}$         &-30.52$^{\ast \ast}$       &-81.06$^{\ast}$            &-31.60$^{\ast}$             &50.88                          \\
$\beta_3$   &-50.98                       &-282.0.0$^{\ast\ast}$     &-117.3$^{\ast \ast \ast}$   &580.7$^{\ast \ast}$         &-40.15                        &-390.4                          &-120.0$^{\ast}$             &409.2                          \\  
$\beta_4$   &                                 &269.6                         &                                     &-62.44                            &                                    &-31.13                      &                                  &-75.60                       \\
$\beta_5$   &                                &-226.2                         &                                      &-147.0 $^{\ast \ast}$      &                                   &3.656                          &                                   &-83.45                      \\
$\beta_6$   &                                 &426.6                          &                                   &-899.0$^{\ast \ast \ast}$  &                                   &652.9$^{\ast}$               &                                &-660.0$^{\ast}$            \\  
Return parameters&                        &                                 &           &             &           &           &               &                  \\
$\delta_1$  &                                 &                                 &1.773                             &1.561                              &                                 &                                   & 2.633$^{\ast \ast}$        &2.140        \\
$\delta_2$  &                                 &                                 &-7.797$^{\ast \ast \ast}$   &-8.785$^{\ast \ast \ast}$  &                                 &                                  & -5.748$^{\ast}$              &-6.396$^{\ast \ast}$   \\  
$\delta_3$  &                                 &                                 &16.10$^{\ast \ast}$           &12.59$^{\ast}$                  &                                &                                   & 7.968                           &11.01$^{\ast}$         \\
$\delta_4$  &                                 &                                 &9.030$^{\ast \ast \ast}$    &9.084$^{\ast \ast \ast}$     &                                &                                  & 8.208$^{\ast \ast \ast}$  & 8.727$^{\ast \ast \ast}$ \\
$\delta_5$  &                                 &                                 &31.45$^{\ast \ast \ast}$    &35.31$^{\ast \ast \ast}$     &                                &                                  & 29.13$^{\ast \ast \ast}$  & 31.25$^{\ast \ast \ast}$ \\
$\delta_6$  &                                 &                                 &13.30$^{\ast }$                 &23.88$^{\ast \ast \ast}$    &                                 &                                  & 24.40$^{\ast \ast \ast}$  & 21.71$^{\ast \ast \ast}$ \\  
                &                                 &                                 &           &             &           &           &               &                  \\
Adj.$R^2$   & 0.564                         & 0.568                        &0.610                              &0.611                              &   0.590                       &  0.591                        & 0.621                           &  0.621                \\
    \hline 
   \end{tabular}
\end{center}
$^{***}$, $^{**}$, and $^{*}$ denote rejections of null hypothesis at 1\%, 5\%, and 10\% significance levels, respectively. Adj.$R^2$ denotes the adjusted $R^2$. 
\end{landscape}

\newpage
\fontsize{11pt}{14pt}\selectfont
\begin{landscape}
\begin{center}
Table 4: Estimation results for the spot market of Nikkei 225
\end{center}

\begin{center}
\begin{tabular}{c|cccc|cccc} \hline       
                 &HAR-J                         &HAR-AJ                    &HAR-J-LE                     & HAR-AJ-LE                &RSV-J                          &RSV-AJ                       &RSV-J-LE                       & RSV-AJ-LE           \\ \hline 
constant     &                                   &                                  &                                  &                                   &                                   &                                 &                                   &                              \\
$c$            &0.049$^{\ast \ast \ast}$&-0.380                        &-0.649$^{\ast \ast \ast}$   &-0.997$^{\ast \ast \ast}$&0.733$^{\ast \ast\ast }$   &0.266                          &-0.116                          &-0.383$^{\ast}$          \\
RV parameters&           &           &           &            &            &            &              &                 \\
$\alpha_1$  &0.386$^{\ast \ast \ast}$ &0.384$^{\ast \ast \ast}$&0.302$^{\ast \ast \ast}$   &0.312$^{\ast \ast \ast}$   &0.110$^{\ast \ast \ast}$  &0.105$^{\ast \ast \ast}$& 0.110$^{\ast \ast \ast}$& 0.110$^{\ast \ast \ast}$\\
$\alpha_2$  &0.385$^{\ast \ast \ast}$ &0.393$^{\ast \ast \ast}$&0.365$^{\ast \ast \ast}$   &0.365$^{\ast \ast \ast}$   &0.086$^{\ast \ast}$         &0.111$^{\ast \ast}$       & 0.145$^{\ast \ast \ast}$ &0.139$^{\ast \ast \ast}$\\
$\alpha_3$  &0.230$^{\ast \ast \ast}$ &0.185$^{\ast \ast \ast}$&0.274$^{\ast \ast \ast}$   &0.234$^{\ast \ast \ast}$   &0.113$^{\ast \ast}$         &0.199$^{\ast }$            & 0.167$^{\ast \ast \ast}$&0.165$^{\ast }$               \\  
$\alpha_4$  &                                 &                                  &                                   &                                    &0.252$^{\ast \ast \ast}$  &0.245$^{\ast \ast \ast}$& 0.199$^{\ast \ast \ast}$ &0.199$^{\ast \ast \ast}$\\
$\alpha_5$  &                                 &                                  &                                   &                                     &0.292$^{\ast \ast \ast}$  &0.284$^{\ast \ast \ast}$& 0.200$^{\ast \ast \ast}$& 0.212$^{\ast \ast \ast}$\\
$\alpha_6$  &                                 &                                  &                                   &                                    &0.144$^{\ast \ast }$         &0.014                         & 0.107$^{\ast}$              & 0.081                          \\  
Jump parameters&                           &           &           &            &            &            &              &                 \\
$\beta_1$   &-238.2                         &-1325.4$^{\ast\ast\ast}$&-449.9                            &-944.6$^{\ast}$             &8.811                            &459.9                         &-320.6                         &-311.3                      \\
$\beta_2$   &-3065.8$^{\ast\ast\ast}$&-3177.6$^{\ast\ast}     $&-3402.5$^{\ast \ast \ast}$&-1771.5                        &-2806.6$^{\ast \ast}$       &-2184.8$^{\ast}$        &-3011.2$^{\ast\ast\ast}$&-1499.2                      \\
$\beta_3$   &-3015.0$^{\ast\ast}$     &333.5                        &-3275.3$^{\ast \ast}$         &1076.3                           &-3410.9$^{\ast \ast}$     &-2007.9                        &-3234.3$^{\ast\ast}$    &-150.1                       \\  
$\beta_4$   &                                 &378.6                         &                                     &-728.4                            &                                    &547.5$^{\ast}$              &                                  &-580.9                       \\
$\beta_5$   &                                &-750.2                         &                                      &-2378.5$^{\ast \ast \ast}$&                                   &-1586.8$^{\ast}$            &                               &-2451.6$^{\ast \ast \ast}$\\
$\beta_6$   &                                 &994.8                          &                                   &-1338.1                            &                                   &1728.9$^{\ast}$               &                                &-1083.5            \\  
Return parameters&                        &                                 &           &             &           &           &               &                  \\
$\delta_1$  &                                 &                                 &1.079                              &3.295$^{\ast \ast }$         &                                 &                                   & 4.423$^{\ast \ast \ast}$ &5.711$^{\ast \ast \ast}$  \\
$\delta_2$  &                                 &                                 &-1.376                            &-0.437                             &                                 &                                  &  0.598                           &1.706                          \\  
$\delta_3$  &                                 &                                 &15.70$^{\ast \ast}$           &16.98$^{\ast \ast }$          &                                &                                   & 13.20$^{\ast\ast}$          &15.10$^{\ast\ast}$      \\
$\delta_4$  &                                 &                                 &7.097$^{\ast \ast \ast}$    &6.631$^{\ast \ast \ast}$     &                                &                                  & 3.705$^{\ast \ast}$         & 4.221$^{\ast\ast}$ \\
$\delta_5$  &                                 &                                 &16.14$^{\ast \ast \ast}$    &17.54$^{\ast \ast \ast}$     &                                &                                  & 14.17$^{\ast \ast \ast}$  & 15.43$^{\ast \ast \ast}$ \\
$\delta_6$  &                                 &                                 &2.901                              &9.297                              &                                 &                                  & 4.087                           & 8.604                          \\  
                &                                 &                                 &           &             &           &           &               &                  \\
Adj.$R^2$   & 0.669                         & 0.670                        &0.684                              &0.684                              &   0.674                       &  0.673                        & 0.685                            &  0.685                \\
    \hline 
   \end{tabular}
\end{center}
$^{***}$, $^{**}$, and $^{*}$ denote rejections of null hypothesis at 1\%, 5\%, and 10\% significance levels, respectively. Adj.$R^2$ denotes the adjusted $R^2$.
\end{landscape}

\newpage
\fontsize{11pt}{14pt}\selectfont
\begin{landscape}
\begin{center}
Table 5: Results of Diebold-Mariano tests in the futures market of Nikkei 225
\end{center}

\begin{center}
\begin{tabular}{c|c|ccccccc} \hline      
                         &                    &\multicolumn{7}{c}{Comparison model} \\ \hline 
Benchmark  model&loss function   &HAR-AJ     &HAR-J-LE                     & HAR-AJ-LE                &RSV-J                       &RSV-AJ                   &RSV-J-LE                     & RSV-AJ-LE     \\ \hline 
HAR-J                &  MSE            & -1.167        &2.968$^{\ast \ast \ast}$ &-0.647                        & 0.859                        &-1.248                        & 1.526                         &-0.826             \\
                         &  MAE            & -1.045        &8.445$^{\ast \ast \ast}$ &4.698$^{\ast \ast \ast}$&6.361$^{\ast \ast \ast}$& 0.575                        & 9.684$^{\ast \ast \ast}$&4.113$^{\ast \ast \ast}$             \\
                         &  HMSE          & -1.300        &0.890                           &-1.056                        &-0.351                        &-1.590                        &-1.390                        &-1.176            \\
                         &  HMAE          & -1.174        &6.724$^{\ast \ast \ast}$ &3.385$^{\ast \ast \ast}$ &4.824$^{\ast \ast \ast}$&-0.082                        &7.076$^{\ast \ast \ast}$&2.814$^{\ast \ast \ast}$            \\  \hline
HAR-AJ            &  MSE            &                    &1.313                          &1.309                            &1.207                        &1.090                          &1.287                         &1.379            \\
                        &  MAE            &                    &4.014$^{\ast \ast \ast}$&5.570$^{\ast \ast \ast}$&2.527$^{\ast \ast }$       &3.030$^{\ast \ast \ast}$ &4.636$^{\ast \ast \ast}$&7.120$^{\ast \ast \ast}$             \\
                        &  HMSE          &                    &1.347                         &1.146                              &1.196                       &0.911                           &1.187                         &1.030            \\
                        &  HMAE          &                    &3.254$^{\ast \ast \ast}$&4.230$^{\ast \ast \ast}$  &2.149$^{\ast \ast \ast}$&2.423$^{\ast \ast}$       &3.602$^{\ast \ast \ast}$&5.167$^{\ast \ast \ast}$            \\  \hline 
HAR-J-LE          &  MSE             &                    &                                 &-1.268                        &-3.212$^{\ast \ast \ast}$&-1.607                          &-0.468                        &-1.176            \\
                       &  MAE            &                    &                                 &-1.265                         &-5.557$^{\ast \ast \ast}$&-4.041$^{\ast \ast \ast}$&4.693$^{\ast \ast \ast}$&-0.629            \\
                       &  HMSE          &                    &                                 &-1.544                         &-1.839$^{\ast}$              &-1.730$^{ \ast}$             &-0.901                       &-1.451             \\
                       &  HMAE          &                    &                                 &-1.318                        &-4.917$^{\ast \ast \ast}$ &-3.341$^{\ast \ast \ast}$&2.704$^{\ast \ast \ast}$&-0.877            \\  \hline
HAR-AJ-LE      &  MSE            &                    &                                   &                                &0.872                            &-1.850$^{\ast}$               &1.207                        &-1.047             \\
                       &  MAE            &                    &                                 &                                 &-2.124$^{\ast \ast}$        &-5.728$^{\ast \ast \ast}$ &2.558$^{\ast \ast}$      &1.029             \\
                       &  HMSE          &                    &                                  &                                & 1.147                           &-1.386                           &1.257                         &-1.260             \\
                       &  HMAE          &                    &                                 &                                 &-1.474                           &-4.632$^{\ast \ast \ast}$ &2.171$^{\ast \ast}$      &0.393           \\  \hline 
RSV-J              &  MSE            &                    &                                 &                                 &                                   &-1.363                           &2.633$^{\ast \ast \ast}$&-0.949             \\
                       &  MAE            &                    &                                 &                                 &                                   &-1.852$^{\ast}$                &7.963$^{\ast \ast \ast}$&2.018$^{\ast \ast}$            \\
                       &  HMSE          &                    &                                 &                                 &                                  &-1.490                             &0.296                          &-1.198            \\
                       &  HMAE          &                    &                                &                                  &                                  & -1.737$^{\ast}$                &6.512$^{\ast \ast \ast}$&1.264            \\  \hline 
RSV-AJ           &  MSE            &                    &                                &                                  &                                  &                                     &1.559                           & 0.645           \\
                       &  MAE            &                    &                                &                                  &                                  &                                     &5.045$^{\ast \ast \ast}$  & 7.060$^{\ast \ast \ast}$            \\
                       &  HMSE          &                    &                                &                                  &                                  &                                     &1.475                            & 0.469             \\
                       &  HMAE          &                    &                                &                                  &                                  &                                     &3.943$^{\ast \ast \ast}$  &5.423$^{\ast \ast \ast}$            \\  \hline
RSV-J-LE       &  MSE            &                    &                                &                                  &                                  &                                     &                                      &-1.139             \\
                       &  MAE            &                    &                                &                                  &                                  &                                     &                                   &-1.607             \\
                       &  HMSE          &                    &                                &                                  &                                  &                                     &                                    &-1.270             \\
                       &  HMAE          &                    &                                &                                  &                                  &                                     &                                   &-1.513             \\  \hline \\
   \end{tabular}
\end{center}
$^{***}$, $^{**}$, and $^{*}$ denote rejections of null hypothesis at 1\%, 5\%, and 10\% significance levels, respectively.
\end{landscape}

\newpage
\fontsize{11pt}{14pt}\selectfont
\begin{landscape}
\begin{center}
Table 6: Results of Diebold-Mariano tests in the spot market of Nikkei 225
\end{center}

\begin{center}
\begin{tabular}{c|c|ccccccc} \hline      
                         &                    &\multicolumn{7}{c}{Comparison model} \\ \hline 
Benchmark  model&loss function  &HAR-AJ      &HAR-J-LE                    & HAR-AJ-LE                &RSV-J                       &RSV-AJ                    &RSV-J-LE                    & RSV-AJ-LE     \\ \hline 
HAR-J                &  MSE   & 0.732                  &3.886$^{\ast \ast \ast}$ &4.770$^{\ast \ast \ast}$ & 2.277$^{\ast \ast}$     &1.141                        &4.145$^{\ast \ast \ast}$ &4.760$^{\ast \ast \ast}$             \\
                        &  MAE    & 2.248$^{\ast \ast}$&5.268$^{\ast \ast \ast}$ &4.946$^{\ast \ast \ast}$&3.966$^{\ast \ast \ast}$&2.379$^{\ast \ast}$     &5.205$^{\ast \ast \ast}$&4.706$^{\ast \ast \ast}$             \\
                         &  HMSE &0.284                    &2.747$^{\ast \ast \ast}$ &3.509$^{\ast \ast \ast}$&2.788$^{\ast \ast \ast}$&1.130                        &3.123$^{\ast \ast \ast}$&3.816$^{\ast \ast \ast}$            \\
                         &  HMAE & 1.867$^{\ast}$       &4.827$^{\ast \ast \ast}$ &4.622$^{\ast \ast \ast}$ &4.277$^{\ast \ast \ast}$&2.260$^{\ast \ast}$     &4.885$^{\ast \ast \ast}$&4.508$^{\ast \ast \ast}$            \\  \hline
HAR-AJ             &  MSE            &                    &3.400$^{\ast \ast \ast}$&4.632$^{\ast \ast \ast}$&0.882                        &0.427                          &3.651$^{\ast \ast \ast}$ &4.292$^{\ast \ast \ast}$            \\
                        &  MAE            &                    &3.663$^{\ast \ast \ast}$&3.975$^{\ast \ast \ast}$&1.198                        &0.573                        &3.853$^{\ast \ast \ast}$&3.596$^{\ast \ast \ast}$             \\
                        &  HMSE          &                    &2.540$^{\ast \ast}$      &3.838$^{\ast \ast \ast}$ &1.101                       &0.962                           &2.795$^{\ast \ast \ast}$&3.624$^{\ast \ast \ast}$            \\
                        &  HMAE          &                    &3.524$^{\ast \ast \ast}$&3.901$^{\ast \ast \ast}$  &1.292                      &0.789                          &3.728$^{\ast \ast \ast}$&3.574$^{\ast \ast \ast}$            \\  \hline 
HAR-J-LE        &  MSE             &                    &                                 &0.810                        &-2.240$^{ \ast \ast}$     &-3.195$^{\ast \ast \ast}$&0.282                        &0.694            \\
                       &  MAE            &                    &                                 &0.018                         &-2.587$^{\ast \ast \ast}$&-3.239$^{\ast \ast \ast}$&0.435                       &-0.158            \\
                       &  HMSE          &                    &                                 &0.657                         &-1.369                         &-2.202$^{ \ast \ast}$       &0.846                       &0.708             \\
                       &  HMAE          &                    &                                 &-0.020                        &-2.343$^{ \ast \ast}$    &-3.102$^{\ast \ast \ast}$&0.509                         &-0.163            \\  \hline
HAR-AJ-LE      &  MSE            &                    &                                   &                                &-3.922$^{\ast \ast \ast}$&-4.596$^{\ast \ast \ast}$ &-0.823                      &-0.463             \\
                       &  MAE            &                    &                                 &                                 &-2.647$^{\ast \ast \ast}$&-3.465$^{\ast \ast \ast}$ &0.229                        &-0.353           \\
                       &  HMSE          &                    &                                  &                                &-2.495$^{\ast \ast}$       &-3.489$^{\ast \ast \ast}$ &-0.505                         &0.094             \\
                       &  HMAE          &                    &                                 &                                 &-2.439$^{\ast \ast}$       &-3.361$^{\ast \ast \ast}$ &0.262                         &-0.329           \\  \hline 
RSV-J              &  MSE            &                    &                                 &                                 &                                   &-0.904                           &2.668$^{\ast \ast \ast}$&4.044$^{\ast \ast \ast}$             \\
                       &  MAE            &                    &                                 &                                 &                                   &-1.087                          &3.038$^{\ast \ast \ast}$&2.668$^{\ast \ast}$            \\
                       &  HMSE          &                    &                                 &                                 &                                  &-0.880                             &1.738$^{\ast}$            &2.803$^{\ast \ast \ast}$            \\
                       &  HMAE          &                    &                                &                                  &                                  & -1.076                           &2.724$^{\ast \ast \ast}$&2.463$^{\ast \ast}$            \\  \hline 
RSV-AJ           &  MSE            &                    &                                &                                  &                                  &                                     &3.709$^{\ast \ast \ast}$ &4.822$^{\ast \ast \ast}$           \\
                       &  MAE            &                    &                                &                                  &                                  &                                     &3.741$^{\ast \ast \ast}$  & 3.585$^{\ast \ast \ast}$            \\
                       &  HMSE          &                    &                                &                                  &                                  &                                     &2.635$^{\ast \ast \ast}$  & 3.824$^{\ast \ast \ast}$             \\
                       &  HMAE          &                    &                                &                                  &                                  &                                     &3.537$^{\ast \ast \ast}$  &3.462$^{\ast \ast \ast}$            \\  \hline
RSV-J-LE        &  MSE            &                    &                                &                                  &                                  &                                     &                                      &0.733             \\
                       &  MAE            &                    &                                &                                  &                                  &                                     &                                   &-0.481             \\
                       &  HMSE          &                    &                                &                                  &                                  &                                     &                                    &0.577             \\
                       &  HMAE          &                    &                                &                                  &                                  &                                     &                                   &-0.459             \\  \hline 
   \end{tabular}
\end{center}
$^{***}$, $^{**}$, and $^{*}$ denote rejections of null hypothesis at 1\%, 5\%, and 10\% significance levels, respectively.
\end{landscape}

\newpage
\fontsize{10pt}{10pt}\selectfont
\begin{figure}
\begin{center}
\caption{Time series plot of $RV_t$ for futures market of Nikkei 225}
   \includegraphics[width=15cm,height=8cm,clip]{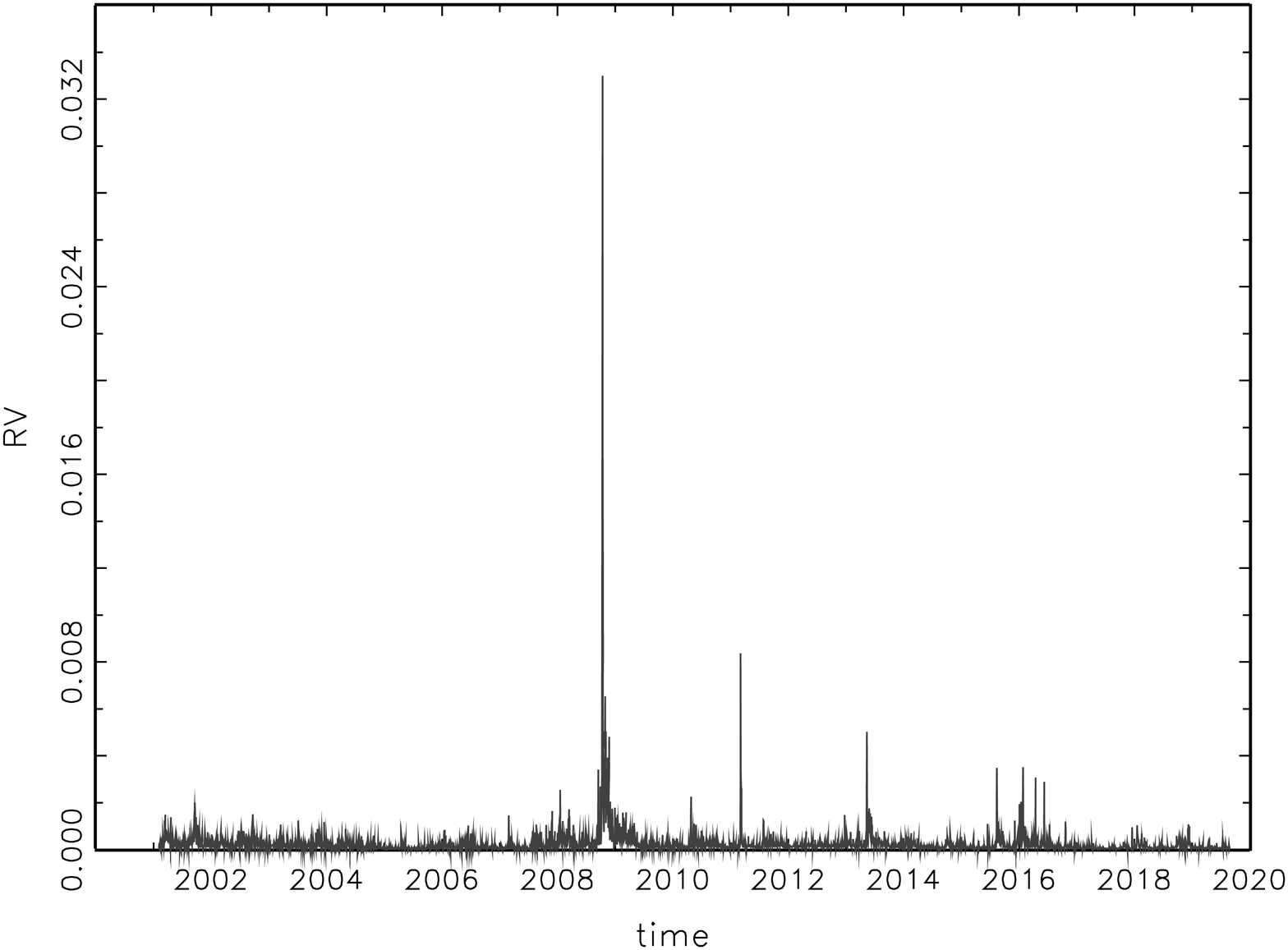} \\
\caption{Time series plot of $\ln RV_t$ for futures market of Nikkei 225}
   \includegraphics[width=15cm,height=8cm,clip]{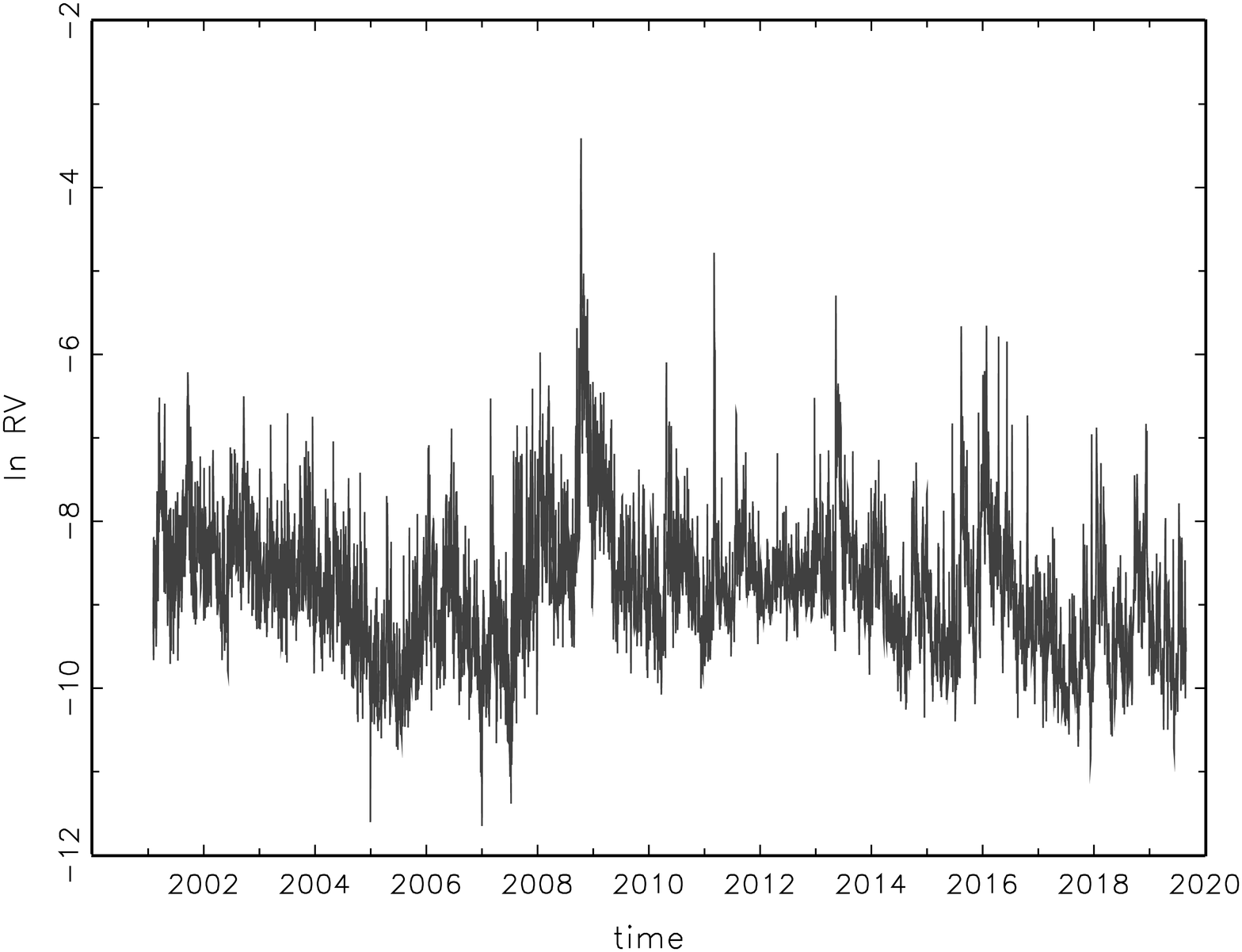} \\
\end{center}
\end{figure}

\newpage
\fontsize{10pt}{10pt}\selectfont
\begin{figure}
\begin{center}
\caption{Time series plot of $RV_t$ for spot market of Nikkei 225}
   \includegraphics[width=15cm,height=8cm,clip]{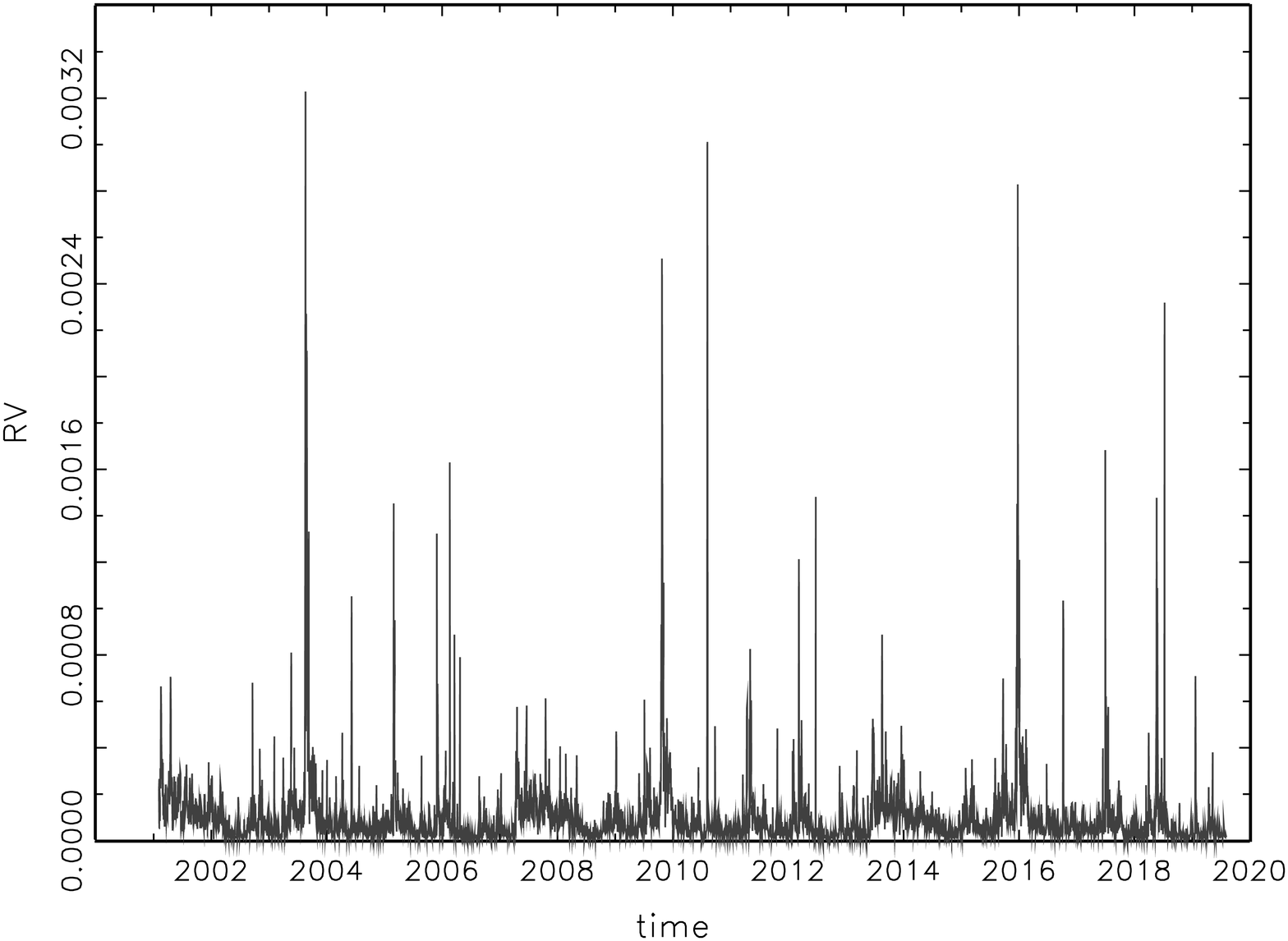} \\
\caption{Time series plot of $\ln RV_t$ for spot market of Nikkei 225}
   \includegraphics[width=15cm,height=8cm,clip]{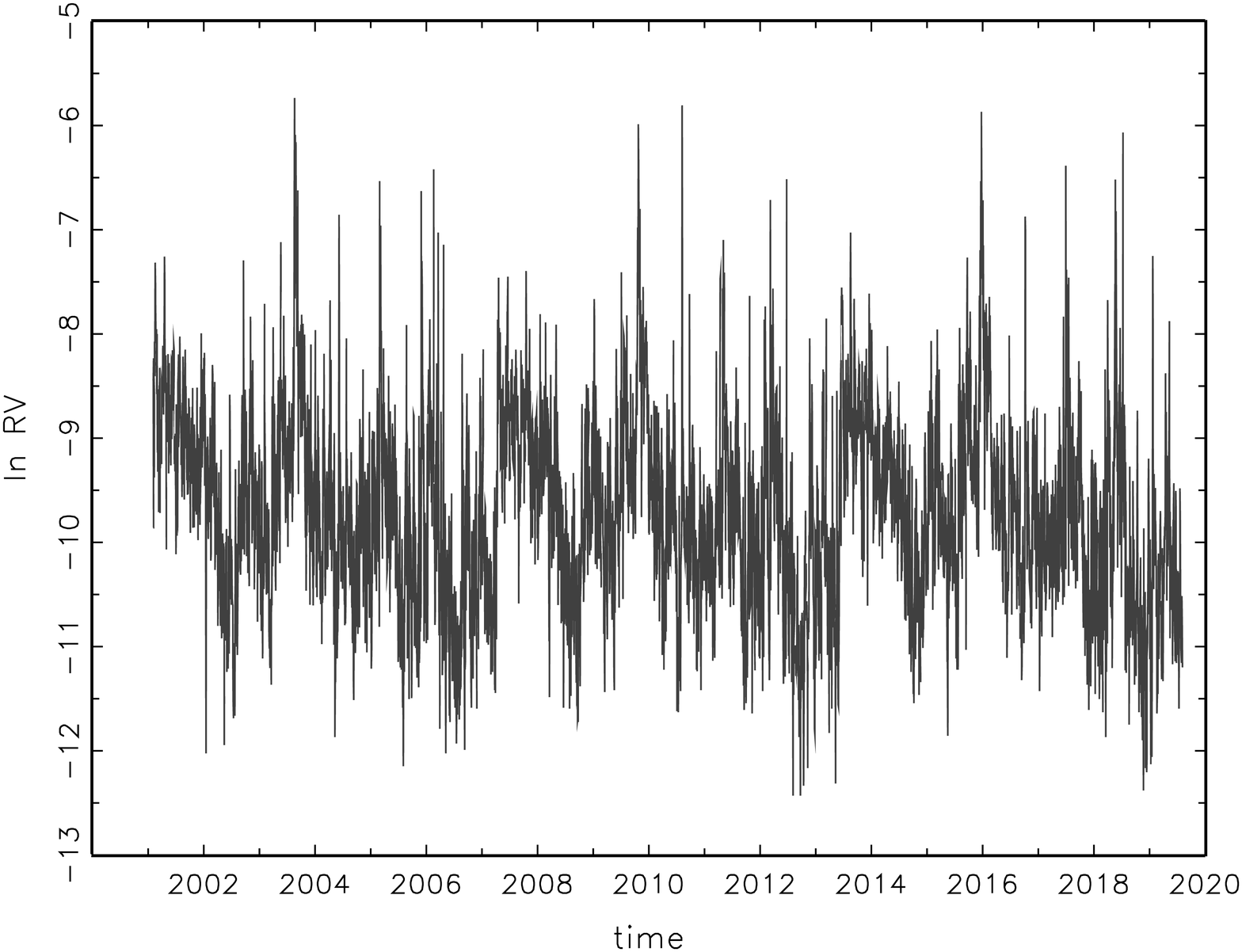} \\
\end{center}
\end{figure}

\end{document}